\documentclass{article}
{}
{}
{}

\pdfoutput=1
\usepackage{amsmath} 
\usepackage{amsthm} 

\usepackage{hyperref} 
\usepackage{graphicx}
\usepackage{algorithmic} 
\usepackage{hyperref}
\usepackage{authblk}
\usepackage[margin=1.2cm]{geometry}
\usepackage{multicol}
\usepackage{blindtext}
\usepackage{epstopdf}
\usepackage{caption}
\usepackage{titlesec}
\usepackage{biblatex}
\usepackage{lipsum}
\titlespacing*{\section}{0pt}{0.1\baselineskip}{0.2\baselineskip}
\begin{document}

\title{\textbf{Redactable Signature Schemes and Zero-knowledge Proofs: A comparative examination for applications in Decentralized Digital Identity Systems}}

\author{Bryan Kumara}
\author{Mark Hooper}
\author{Carsten Maple}
\author{Timothy Hobson}
\author{Jon Crowcroft}
\affil{{The Alan Turing Institute, London, U.K.} Email: bkumara, mhooper, cmaple, thobson, jcrowcroft @turing.ac.uk}
   
\date{23 July 2023}

\renewcommand\Affilfont{\itshape\small}

\maketitle

\begin{abstract}
Redactable Signature Schemes and Zero-Knowledge Proofs are 
two radically different approaches to enable privacy. This paper analyses their merits and drawbacks when applied to decentralized identity system. Redactable Signatures, though competitively quick and compact, are not as expressive as zero-knowledge proofs and do not provide the same level of privacy. On the other hand, zero-knowledge proofs can be much faster but some protocols require a trusted set-up. We conclude that given the benefits and drawbacks, redactable signatures are more appropriate at an earlier stage and zero-knowledge proofs are more appropriate at a later stage for decentralized identity systems.

\end{abstract}

\begin{multicols}{2}

\section{Introduction}
A critical aspect of economic growth is the integration of the internet as a medium for innovation and exchange \cite{Ali:2013}. It is vital that digital identities which computationally represent external agents are able to address stakeholder concerns as each digital identity system comes with its own set of trade-offs that considers various factors such as security, reliability, resilience, accessibility, and scalability \cite{digital} \cite{Ali:2013} \cite{Camp:2004} \cite{Atick:2016}.
A key issue these systems may consider is the degree to which a single party may exert control over the operation of the system. Centralized digital identity systems usually contain a dedicated registration process and store information in a central database \cite{Fang:2009}. However, the centralization of authority and storage carries the risk of abuse such as exclusion and censorship alongside individual profiling \cite{Ellemers:2002}. 

One approach to address these concerns is to adopt a decentralized identity system. This approach typically involves the use of transparent, open networks and protocols with voluntary participation in the attempt to minimize data collection or formal enrolment \cite{Dib:2020}. To achieve this, Decentralized Identifiers (DIDs) and Verifiable Credentials (VC) are typically used. The former is a globally unique identifier which allows a party to be identified in a verifiable and persistent manner without the need of a centralized registry. The latter is an open standard for digital credentials with a digital signature. A VC can be issued by anyone and can be presented for verification by everyone. A combination of DIDs and VCs gives individual parties control of their digital attributes without a central authority \cite{Lux:2020}.

DIDs and VCs rely on some verification mechanism to establish truth. This can be done through cryptographic protocols such as a digital signature which ensures a message was created by a known sender without tampering \cite{kaur:2012}. There are interesting protocols one can construct using digital signatures, and in this paper we will discuss Redactable Signature Scheme (RSS). RSS builds on top of digital signatures by allowing a third party to redact or remove parts of a message and still derive a valid signature. Another approach is to adopt Zero-Knowledge Proof (ZKP), which is a mathematical argument that convinces verifiers that some computation has been performed correctly but reveals nothing else. ZKPs are complex constructions, and can be implemented with variants that provide their own set of trade-offs. 

Both RSS and ZKPs address the issue of privacy-preserving verification in the context of decentralized identity systems. Individual identities no longer need to pass through a centralized system where empowered nodes could abuse their privilege, but users can also protect their privacy when presenting their digital identity by retaining agency over what information they would like to disclose to others. RSS allows one to validly redact parts of a VC, whilst ZKPs allows users to prove that they meet some access criteria without having to show their VC at all. These protocols address the issue of privacy and agency for credential holders. For instance, a decentralized identity system may be used to check that a user is not below the legal age limit for purchasing alcohol when entering a bar. The credential holder would present their VC to a verifier for a check on their age, which is visible in the attributes section. However, a holder may not want to present the whole of their credential, which may include other attributes such as address, marital status, and nationality. RSS allows the redaction of other attributes, leaving only the birth date on the VC for verification. ZKPs takes it a step further, with the verifiers accepting a proof that the holder of a VC is older than the legal age but never seeing the VC itself. 

This paper examines the benefits and drawbacks of RSS and ZKPs for VCs. Though both schemes enable a higher degree of privacy and control for the holder of a credential, they are very different protocols that impacts the overall system in various ways. We conclude that given their trade-offs, they can actually be harmonized by being implemented at different stages of an decentralized identity system to provide a high level of user privacy to best ameliorate their shortcomings.

\section{Background}
\subsection{Redactable Signature Schemes}
Public-key cryptography, which encompasses digital signatures and zero-knowledge proofs, rely on the hardness of solving some mathematical problems. This paper focuses pairing-based cryptographic techniques, which consists of three bilinear groups $\mathbf{G}_1, \mathbf{G}_2, \mathbf{G}_T$ and a bilinear map $e: \mathbf{G}_1 \times \mathbf{G}_2 \rightarrow \mathbf{G}_T$. These groups and mappings satisfies the following properties for some prime order $p$:
\begin{itemize}
    \item for all $g \in \mathbf{G}_1, \tilde{g}\in \mathbf{G}_2$ and $a,b\in \mathbf{Z}_p$, $e(g^a,\tilde{g}^b)=e(g,\tilde{g})^{a\dot b}$;
    \item for $g \neq 1_{\mathbf{G}_1}$ and $\tilde{g} \neq 1_{\mathbf{G}_2}$, $e(g,\tilde{g})\neq 1_{\mathbf{G}_T}$;
    \item the map $e$ is efficiently computable.
\end{itemize}

There are three different types of pairings - Type 1 pairing where $\mathbf{G}_1 = \mathbf{G}_2$, Type 2 pairing where $\mathbf{G}_1 \neq \mathbf{G}_2$ with an efficiently computable homomorphism $\phi': \mathbf{G}_2 \rightarrow \mathbf{G}_1$, and Type 3 pairing where $\mathbf{G}_1 \neq \mathbf{G}_2$ with no efficiently computable homomorphishms $\phi:\mathbf{G}_1 \rightarrow \mathbf{G}_2$ and $\phi': \mathbf{G}_2 \rightarrow \mathbf{G}_1$. We only consider Type 3 pairing in this paper \cite{Sanders:2016}.

The RSS construction we examine relies on following computational assumptions for its security \cite{Sanders:2016}:
\begin{itemize}
    \item Symmetric Discrete Logarithm (SDL) Assumption: Given $(g,g^a)\in \mathbf{G}_1^2$ and $(\tilde{g},\tilde{g}^a)\in \mathbf{G}_2^2$, it is hard to recover $a$.
    \item Extended Decisional Diffie-Hellman (EDDH) Assumption: Given $(\{g^{a \cdot c^i}\}_{i=0}^{2n^2}, \{g^{b\cdot c^i}\}_{i=0}^{n-1}, \{g^{c^i}\}_{i=1}^{3n^2}, \{g^{d\cdot c^i}\}_{i=1}^{2n^2}\in \mathbf{G}_1^{7n^2+n+2}$ and $(\{\tilde{g}^{c^i}\}_{i=1}^{2n^2}, \tilde{g}^d, \{\tilde{g}^{a \cdot c^1}\}_{i\in[1,2n^2]\setminus[n^2-n,n^2+n]}$, it is hard to decide if $z=a\cdot b \cdot c^{n^2}+b \cdot d$ or $z$ is random.
\end{itemize}

With these foundations, we can construct the Pointcheval-Sanders (PS) digital signature. This signature is re-randomizable, meaning that given a valid signature one can derive a new version of that signature on the same message by introducing a random element \cite{Sanders:2021}. It will be a useful property as we construct RSS from PS. To sign a message $m=(m_1,...m_r)\in \mathbf{Z}_p^r$, the PS scheme does the following \cite{Sanders:2016}:
\begin{itemize}
    \item $Setup(1^k)$: Given a security parameter $k$, output $pp \leftarrow (p,\mathbf{G}_1,\mathbf{G}_2,\mathbf{G}_T,e)$ of type 3 bilinear groups. 
    \item $Keygen(pp)$: given the public parameters, $\tilde{g} \overset{{\scriptscriptstyle\$}}{\leftarrow} \mathbf{G}_2$  and $(x,y_1,...,y_r) \overset{{\scriptscriptstyle\$}}{\leftarrow} \mathbf{Z}_p^{r+1}$, compute $(\tilde{X},\tilde{Y}_1,...,\tilde{Y}_r)\leftarrow (\tilde{g}^x,\tilde{g}^{y_1},...,\tilde{g}^{y_r})$. Let the secret key $sk=(x,y_1,...,y_r)$ and the public key $pk=(\tilde{g},\tilde{X},\tilde{Y}_1,...,\tilde{Y}_r)$.
    \item $Sign(sk,m_1,...,m_n)$: Generate a random $h \overset{{\scriptscriptstyle\$}}{\leftarrow} \mathbf{G}_2$ and outputs the signature $\sigma \leftarrow (h,h^{(x+\Sigma y_j \cdot m_j))}$
    \item $Verify(pk,(m_1,...m_r),\sigma)$: Given a signature $\sigma=(\sigma_1,\sigma_2)$, check if $\sigma_1 \neq 1_{\mathbf{G}_1}$ and $e(\sigma_1,\tilde{X}\cdot\Pi \tilde{Y}_j^{m_j})=e(\sigma_2,\tilde{g})$ are satisfied. If it is, return $1$ else return $0$
\end{itemize}
For proofs of correctness and security alongside a more comprehensive overview of the PS signature scheme, readers should refer to \cite{Sanders:2016}. Given the PS signature scheme, one only needs slight modification to reach the RSS construction. The idea of this RSS is that given a set of messages $\{m_i\}_{i\in I}$ and an associated signature $\sigma$, one can redact parts of a message $i\in\overline{I}$ for some $I \subset [1,n]$ and $\overline{I}= [1,n]\setminus I$ such that the signature is valid for $\{m_i\}_{i \in I}$. For a more comprehensive treatment of RSS that covers alternative constructions and features, readers should refer to \cite{bilzhause:2017}. 

Intuitively, this RSS construction uses the PS signature scheme and aggregates the redacted messages into one element. To prove that this element was generated honestly a second element is calculated from the first. To avoid a quadratic scaling based on the length of the message that previous constructions RSS suffered from, this scheme starts from a modified instance of the PS scheme where the secret key contains two scalars $x,y$ and a hash function is used to create scalars that indicate a commitment on which indexes $i\in [1,n]$ have been redacted, preventing a maliciously generated signature. A second pairing equation is added for the verification process to check the scalars generated from the hash function. This results in a constant signature size and a public key that scales linearly to the message length $n$, which is a notable improvement from previous RSS schemes. A full examination of this RSS construction covering security proofs can be found in \cite{Sanders:2021}. 

Given a message of length $n$, the RSS Scheme we consider proceeds as follows: 
\begin{itemize}
    \item $Keygen(1^k,n)$: Given a security parameter $1^k$ and an integer $n$, generate two random elements $(g,\tilde{g})\leftarrow\mathbf{G}_1 \times \mathbf{G}_2$ and scalars $(x,y) \leftarrow \mathbf{Z}_p^2$. Then, compute the following:
    \begin{enumerate}
        \item $\tilde{X} \leftarrow \tilde{g}^x$
        \item $\tilde{Y_i}\leftarrow \tilde{g}^{y_i} \forall 1 \leq i \leq n$
        \item $Y_i \leftarrow g^{y_i} \forall i \in [1,n] \cup [n+2,2n]$
    \end{enumerate}
    The secret key $sk$ is $(x,y)$ and the public key $pk$ is $(H,g,\tilde{g},\{Y_i\}_{i=1}^n,\{Y_i\}_{i=n+2}^2n,\tilde{X},\{\tilde{Y}_i\}_{i=1}^n)$ where $H:\{0,1\}*\rightarrow \mathbf{Z}_p^*$ is the description of the hash function.
    \item $Sign(sk, \{m_i\}_{i=1}^n)$: Given a secret key $sk$ and $n$ messages $m_1,...,m_n$, generate a random element $\sigma_1 \overset{{\scriptscriptstyle\$}}{\leftarrow} \mathbf{G}_1$ and compute $\sigma_2= \sigma_1^{x+\Sigma_{i=1}^n y_i \cdot m_i}$. Let $\sigma_3=1_{\mathbf{G}_1}$ and $\tilde{\sigma}=1_{\mathbf{G}_2}$. Output a signature $\sigma= (\sigma_1,\sigma_2,\sigma_3,\tilde{\sigma})$.
    \item $Derive(pk,\sigma, \{m_i\}_{i=1}^n, \mathcal{I})$: Given an index $\mathcal{I}\subseteq [1,n]$ of message segments we would like to retain, a message $m$, its associated signature $\sigma$, and a public key $pk$, generate two random scalars $(r,t) \leftarrow \mathbf{Z}_p^2$ and compute the following:
    \begin{itemize}
        \item $\sigma_1' \leftarrow \sigma_1^r$
        \item $\sigma_2' \leftarrow \sigma_2^r \cdot (\sigma_1')^t$
        \item $\tilde{\sigma}' \leftarrow \tilde{g}^t \cdot \Pi_{j\in\overline{\mathcal{I}}}\tilde{Y}_j^{m_j}$
    \end{itemize}
    where $\bar{\mathcal{I}}=[1,n] \setminus \mathcal{I}$. For all $i\in \mathcal{I} $, compute the following scalar $c_i\leftarrow H(\sigma_1' || \sigma_2'|| \tilde{\sigma}' || I || i)$ which is used to calculate:
        \begin{itemize}
        \item $\sigma_3' \leftarrow \Pi_{i\in \mathcal{I}}[Y_{n+1-i}^t \cdot \Pi_{j\in\overline{\mathcal{I}}}Y_{n+1-i+j}^{m_j}]^{c_i}$
    \end{itemize}
    The derived signature is $\sigma' = (\sigma_1',\sigma_2',\sigma_3',\tilde{\sigma}')$ for a redacted message $\{m_i\}_{i\in \mathcal{I}}$.
    \item $Verify(pk,\sigma,\{m_i\}_{i\in \mathcal{I}})$: Given a signature $\sigma$, a public key $pk$, and a message $\{m_i\}_{i\in \mathcal{I}}$, if $\sigma_1 = 1_{\mathbf{G}_1}$, return $0$. Otherwise, test if both equations hold before returning $1$:
    \begin{itemize}
        \item $e(\sigma_1,\tilde{X}\cdot\tilde{\sigma}\cdot\Pi_{i\in \mathcal{I}}\tilde{Y}_i{m_i})=e(\sigma_2,\tilde{g})$;
        \item $e(\sigma_3,\tilde{g})=e(\Pi_{i\in \mathcal{I}}Y_{n+1-1}^{c_i},\tilde{\sigma})$;
    \end{itemize}
\end{itemize}

Note that the public key of the issuer is sufficient to generate a derived signature, which allows a holder of a credential redact and derive signatures without informing the issuer which is just the feature we'd like users to have in a decentralized ecosystem.

\subsection{Zero-Knowlege Proofs}
Zero-knowledge proofs (ZKPs) are methods where a party called the prover can prove to another party called the verifier that a given statement is true without revealing any additional information \cite{Oded:1994}. Traditionally, a prover can show knowledge of a certain piece of information by simply revealing it. However, ZKPs allows one to prove possession of certain secret knowledge called a witness which satisfies some statement without revealing the witness itself or any additional information \cite{Oded:1994}. ZKPs have a diverse range of constructions: some require an initial set-up phase, represent computations differently, and some protocols may involve multiple rounds of interaction between the prover and the verifier \cite{Oded:1994}. 

For the purposes of this paper, we consider a type of ZKP called Zero-Knowledge Succinct Non-Interactive Argument of Knowledge (ZK-SNARK) which are efficient in terms of size and verification time (Succinct), does not require multiple rounds of communication between the prover and the verifier (Non-interactive), and provides sound proofs against a polynomial-bounded prover (Arguments of Knowledge) \cite{Petkus:2019}. Theoretically, a protocol is defined as zero-knowledge with regards to a simulator. This is a machine that operates in a theoretical world that is indistinguishable from the real world for a party that attempts to decide if knowledge was leaked. In this theoretical world, the the simulator can compute a proof without anything else but by just knowing the truth of the statement. A protocol is then zero-knowledge if a proof produced by a simulator who does not know the witness is the same as one produced by a real-world prover that knows the witness \cite{Petkus:2019}. 

The efficiency of ZK-SNARKs in terms of size, speed, and communication makes this an attractive protocol for anonymous disclosure. However, ZK-SNARKs require an initial set-up phase that requires trust. This procedure generates a common set of parameters called the common reference string (CRS) that must be used every time we run a ZK-SNARK proof. CRS generation requires trust as parties involved must delete by-products of the procedure called toxic waste which may otherwise be used to generate forged proofs \cite{Petkus:2019}.  Formally, a ZK-SNARK uses some efficiently decidable binary relation $\mathcal{R}$ with pairs $(u,w)\in\mathcal{R}$ where $u$ is called the statement and $w$ is the witness. A non-interactive argument for $\mathcal{R}$ consists of probabilistic polynomial algorithms $\Pi = (G,P,Ver,Sim)$ where:
\begin{itemize}
    \item $(crs,vrs,td) \leftarrow G(1^\lambda,\mathcal{R})$: Given a securty parameter $\lambda$ and a relation $\mathcal{R}$, output a common reference string $crs$, a trapdoor $td$, and a verification key $vrs$.
    \item $\pi \leftarrow P(crs,u,w)$: Given a common reference string $crs$, a statement $u$, and a witness $w$, the prover algorithm outputs some argument $\pi$.
    \item $b\leftarrow Vers(crs,u,\pi)$: Given an input statement $u$, and argument $\pi$, and a common reference string $crs$. the verifier algorithm outputs $b=1$ if the proof is accepted and $b=0$ otherwise.
    \item $\pi \leftarrow Sim(crs,td,u)$: Given a trapdoor $td$, a statement $u$, and a proof $\pi$, the simulator returns an arguement $\pi$.
\end{itemize}.

An in-depth exploration into the exact procedures of ZK-SNARKS is complicated and lies outside the scope of this paper. \cite{ZKSNARK} \cite{Anca:2020} \cite{Hartwig:2017} \cite{Jo:2019} \cite{Petkus:2019} are particularly good at providing an accessible introduction into this protocol. In short, zk-SNARKs operate by representing computation as a sequence of constrained variables. This is then transformed to polynomials. With the addition of some hiding techniques and using the witness to create a modified set of polynomials, we can map them to pairing-friendly groups. A verifier simply needs to evaluate some pairing functions to check a proof. Informally, ZK-SNARK is able to represent any computation that can be verified quickly, and this allows us to express complex access criteria \cite{Petkus:2019}. There are multiple ways to implement ZK-SNARKs, and this paper considers the Groth16 implementation. This is a general-purpose ZK-SNARK which enables constant size proof and fast verifier time but requires a circuit-specific trusted setup for the pre-processing phase \cite{Groth:2016}. For circuit of arithmetic operations, a Groth16 proof consists of 2 elements in $\mathbf{G}_1$ and 1 element in $\mathbf{G}_2$ which is an improvement from previous implementation such as the Pinocchio ZK-SNARK which contains 7 elements in $\mathbf{G}_1$ and 1 element in $\mathbf{G}_2$ \cite{Groth:2016}. Verification in Groth16 only requires 3 pairings instead of the 12 used in Pinocchio and comes with a shorter CRS \cite{Groth:2016}. In addition to the appealing performance metrics, Groth16 also allows linkage proofs \cite{Rosenberg:2023}. Linkage proofs show that a private set of Groth16 proofs share some subset of hidden public inputs. Verifying a linkage proof does not reveal the Groth16 proof constituents or even the public inputs involved. Linkage proofs are key in constructing the ZKP system we are interested in \cite{Rosenberg:2023}.

zk-Creds is a "flexible, issuer-agnostic anonymous credential toolkit for complex identity statements" \cite{Rosenberg:2023}. The system relies on Groth16 ZK-SNARKs to prove these identity statements, and linkage proofs of Groth16 allows zk-Creds to work with different forms of credentials as long as they have some public credential. This enables post-deployment composition of credentials, and new access criteria. The assumption made by zk-Creds is that there is a list of issued credentials that is either maintained by a trusted party or some decentralized system \cite{Rosenberg:2023}, which is the standard assumption made by VCs.

zk-Creds allows three sets of functionalities: issue, show and verify, and revoke \cite{Rosenberg:2023}. Unlike RSS and VCs in general, zk-Creds credentials does not rely on signatures but rather proof of being in an issued set. In the issuance process, a user requests the issuer that a credential should be issued by meeting some predetermined requirement. Once the credential is placed on the list of issued credential, the user can use it to prove that some access criteria is met to other parties. Using a credential provides a membership proof of the credential being on the list of issued credentials, and a proof that committed attributes of credentials meets the access criteria. A credential can be revoked simply by removing it from the list of issued credentials. In \cite{Rosenberg:2023} notes that there are multiple ways to demonstrate set membership, and uses Merkle forests as a new approach to gain a large reduction in proof generation time but a slight increase in verification time and witness size. An examination of alternative approaches can be found in \cite{Ilker:2021}. 

Groth16 linkage proofs can be combined with membership proof (which shows that an element belongs to a set) and multiple access proofs (which shows that some criteria is fulfilled by a prover) to create Blind Groth16 \cite{Rosenberg:2023}. Blind Groth16 allows us to reuse membership proofs in multiple showings and still maintain privacy \cite{Rosenberg:2023}. This also allows an easy composition of access criteria to prevent determining in advance all the circuits that verifiers will support, prevent the generation of circuit parameters for every combination of access criteria, and avoid dynamical generation of circuit parameters for access criteria as they are needed. Consequently, Blind Groth16s allows one to show that various Groth16 proofs are made based on the same credential without revealing the credential. A full exploration of zk-Creds can be found in \cite{Rosenberg:2023}.

\section{Concrete Performance Metrics}
Previous work on RSS have generally examined its performance with respect to the number of arithmetic operations required, or from a theoretical computational complexity. However, to better assess their suitability for VCs, we would need concrete metrics such as run-time and size in terms of millisecond and bytes respectively. Thus, we have gathered results on both RSS and zk-Creds for VCs when implemented using the Rust programming language. 

The VC we considered is compliant with W3C standards which can be found in \cite{W3C}. This sample VC details a user's personal information such as their name, date of birth, and address (which consists of street name, postcode, and country). For RSS, we want to create a redacted VC which shows that a user resides in a certain country and nothing else. Thus, all the fields but the name the country have been redacted which should give us a good illustration of near-worst-case performance as users redact leave only two fields unredacted. The code for our implementation can be found on \cite{RSSCode}. As this is merely a proof-of-concept, the implementation would be unsuited for immediate deployment in a VC system, and only unit testing has been conducted to ensure that components run as intended.

We collect the speed and size of the RSS algorithm discussed in the previous section. We have 4 algorithms to measure (generating keys, initial signing, deriving a redaction signature, and verifying signatures), gathering the size of the signature, keys, and verification elements, and the results are presented below, rounded to 2 significant figures:
\begin{center}
\begin{tabular}{||c c c||} 
 \hline
 Category & Runtime (ms) & Size (bytes) \\ [0.5ex] 
 \hline\hline
 public key  & 588.22 & 816\\ 
 \hline
 private key  & 7.01 & 112  \\
 \hline
 signature & 545 & 1344  \\
 \hline
 derived signature & 824.84 & 1344 \\
 \hline
 verifying signature & 310.77 & 32 \\ [1ex] 
 \hline
\end{tabular}
\end{center}

For zk-Creds, we use the performance results gathered by \cite{Rosenberg:2023}. Their implementation focused on converting an existing credential into zero knowledge proofs and demonstrating a range of access criteria such as expiry, late limiting, and clone resistance. The results below gives prover-side optimized worst-case cost where the cost of showing a credential is in a membership list is included: 
\begin{center}
\begin{tabular}{||c c c||} 
 \hline
 Client-Opt & Runtime (ms) & Proof Size (bytes) \\ [0.5ex] 
 \hline\hline
 Simple Possession  & 784 & 744\\ 
 \hline
 Expiry  & 875 & 1064  \\
 \hline
 Linkable Show & 879 & 1064  \\
 \hline
 Rate Limiting & 895 & 1064 \\
 \hline
 Clone Resistance & 916 & 1064 \\ 
 \hline
 VerifyShow & 6 & 1064 \\
 \hline
\end{tabular}
\end{center}

The following results are after excluding the cost of the membership list but still prover-side optimized:  
\begin{center}
\begin{tabular}{||c c c||} 
 \hline
 Client-Opt & Runtime (ms) & Proof Size (bytes) \\ [0.5ex] 
 \hline\hline
 Simple Possession  & 5 & 744\\ 
 \hline
 Expiry  & 98 & 1064  \\
 \hline
 Linkable Show & 104 & 1064  \\
 \hline
 Rate Limiting & 117 & 1064 \\
 \hline
 Clone Resistance & 139 & 1064 \\
 \hline
 VerifyShow & 6 & 1064 \\
 \hline
\end{tabular}
\end{center}

Lastly, \cite{Rosenberg:2023} has results where computations are optimized for the verifier. This covers the scenario where a SNARK proof is not reused, and a batch verification is included which indicates the throughout of verifying a set of 100 proofs. All verification numbers are single-threaded and concurrent processing is possible:
\begin{center}
\begin{tabular}{||c c c||} 
 \hline
 Server-Opt & Runtime (ms) & Proof Size (bytes) \\ [0.5ex] 
 \hline\hline
 Simple Possession  & 699 & 744\\ 
 \hline
 Expiry  & 796 & 1064  \\
 \hline
 Linkable Show & 837 & 1064  \\
 \hline
 Rate Limiting & 817 & 1064 \\
 \hline
 Clone Resistance & 812 & 1064 \\ 
 \hline
 VerifyShow & 1.5 & 192 \\
 \hline
 Verify Batch & 1.8 verifs & 192 \\
 \hline
\end{tabular}
\end{center}

\section{Evaluation}
Having gathered concrete metrics from RSS and zk-Creds, we can discuss the two schemes in greater depth for VC applications.

\subsection{RSS for VC}

The RSS construction we considered possesses several attractive properties in theory. Namely, a constant signature size, flexible attribute redaction for selective disclosure, no requirements for a trusted set-up, and possible schemes built on top of it. However, it has some key drawbacks. From the results gathered, the constant signature and private key size of 1344 bytes and 112 bytes respectively are substantially bigger than other signature schemes used in VCs. For instance, the pairing-based BBS+ signature scheme that is in the final stages of standardization has a signature size of 112 bytes and a private key size of 32 bytes alongside a 96 byte public key when instantiated on the widely used BLS12-381 parameters \cite{BBS} \cite{BBSCode}. This is a far more attractive in terms of size, and BBS+ signatures provide proof-of-knowledge schemes that also enables selective disclosure of messages similar to ZK-SNARKs. Given our implementation scenario that considered redacting all but one's name and country on a VC, the BBS+ proof would have a size of 464 bytes. Again, in terms of size, this is a significantly more attractive result than RSS by offering half the size of a derived signature.

Regarding the run-time, most of the RSS algorithms are much slower than BBS+. For instance, generating keys takes 151.81ms for BBS+ but 588.22ms for RSS public keys. RSS key generation has the potential to run much slower too, given that public keys scale linearly with the size of a message. In addition, signing using the BBS+ scheme takes only 57.02ms which is almost 10x faster than the RSS signature and almost 15x faster than generating a derived signature. The only superior performance for RSS is in the verification, where RSS takes 310.77ms to verify whilst verifying a BBS+ proof takes 3478.74ms. This is a roughly 10x difference, but in absolute terms the slowness of BBS+ proof verification will be much more noticeable to users than the comparative difference in BBS+ and RSS signing. The issue of the BBS+ proof verification is also more significant when we consider that the BBS+ proof verification has a linear scaling. Considering a VC with significantly more attributes than our implementation, going beyond 3.5 seconds to verify may become a barrier for adoption.

Though the public key of RSS will scale with the length of the message, it is arguable that given the rare frequency of generating keys this is an acceptable one-off cost unlike the linear growth of a BBS+ proof which has to be used every single time one wants to perform selective disclosure. Thus, if speed is a concern the verification time of RSS may make it preferable than adopting BBS+ signatures.

However, there are other considerations outside of the concrete metrics from RSS. On a general level, RSS offers credential holders a level of flexibility with what information to disclose. Returning to the canonical example of showing that one is above 18 years old, an individual may not want to show any other fields of a VC except for their age. However, other users may have less concerns and be comfortable with displaying their whole or partial VC. RSS allows users to chose their own level of privacy, and this will impact the run-time to generate a derived signature. RSS also does not require a trusted set-up as used in ZK-SNARK proofs. Of course, trust is placed in the issuer of a credential and is a form of centralization, but this form of trust is also placed in a ZK-SNARK anyway where one has to assume that a digital credential is generated by a trustworthy issuer. The lack of a trusted set-up for an RSS scheme means that there is no risk of a forged proof (however small) due to kept toxic waste of a CRS, and a forgery attempt of a redacted signature is thwarted by the security properties of an RSS construction as seen in \cite{Sanders:2021}. 

Lastly, RSS can also be used as a building block for more advanced cryptographic schemes beneficial to VCs. \cite{Sanders:2021} used the RSS implementation considered on this paper to create a group signature (which permits members of a group of anonymously sign a message which represents the group) with a revocation mechanism that enables users to be removed from a group. One of the uses of this feature is for public transport, specifically to implement an anonymous transport subscription pass where users are able to take an unlimited number of trips within a fixed period. Given that VC applications may include that of government registries \cite{Dib:2020}, an anonymous travel pass within a subscription model for public transport would be a perfect fit for VC functionality.

However, there are some more drawbacks for RSS. In addition to the slower and bigger key generation and signature metrics compared to BBS+, validating an RSS requires receivers to perform their own computation on top of the RSS scheme. Though RSS lends itself to convenient constructions as highlighted above in the case of an anonymous transport subscription, it is not as flexible as zk-Creds as we will see in the following sections. Though the public key linear growth is not ideal, it is an improvement over previous quadratic key sizes of the RSS \cite{bilzhause:2017} and again keys do not have to be frequently generated which may be more efficient than the linear growth of BBS+ proof verification scheme. With execution times running at 800ms in the worst case, this is still within the realm of use-ability for DID and VC application. The format of a W3C VC is relatively small with around 6 or so attributes, and unless substantially more attributes would be included it would not drastically impact performance or user experience.

\subsection{zk-Creds for VC}
On the other hand, zk-Creds possess desirable properties for VCs unavailable to RSS. The Groth16 ZK-SNARK has one of the smallest proof size and fastest verification time in ZKP literature with a competitive proof generation \cite{Rosenberg:2023}. Indeed, based on the results from \cite{Rosenberg:2023}, zk-Creds runs with a similar speed when we work with generating a proof alongside its membership witness compared to RSS. For instance, access criteria shown above range from 784ms to 916ms when working with a client-side optimization. In fact, it has a slightly better performance when we work with a server-optimized configuration (though a difference of 100ms would hardly be noticeable in real-life applications). Verifying proofs takes anywhere from 1.8ms to 6ms, is a huge performance improvement over RSS and BBS+. With proof sizes ranging anywhere from 192 bytes to 1064 byes, its clear that the speed and size of zk-Creds is superior when compared to RSS and BBS+. Bear in mind that these results already pass the access criteria check, whereas in RSS a verifier would still need to extract the information from the VC and perform the access check after verifying the redacted signature. The use of sub-circuits called gadgets also ensures that verifiers can create separate proofs that allow pre-computation to further improve speed. Gadgets also allows the representation of more complex access criteria such as session-binding and rate limiting which provides a more fine-grained access control than what is possible with RSS, so there is room for optimization to derive even better performance metrics over RSS .

In addition to performance, Groth16 linkage proofs is an extremely valuable mechanism which allows proofs to show that a hidden collection of Groth16 proofs share hidden public inputs. Given the VC assumption of having a publicly verifiable identity, one could chain different credentials (for instance, a digital driving license, digital work visa, digital marriage certificate) the same way as physical identity documents for an individual all refer to the same person. Possible extension of linkage proofs can be applied internationally between countries to acknowledge travel documents issued by another authority. zk-Creds has shown that it is possible to utilize existing government-issued credentials such as an NFC-enabled passport into a digital credential which is hugely beneficial in easing adoption. This would not be possible with RSS, as different credentials would have to be redacted separately and sent for verification or verifiers must be re-configured to accept different verification methods. However, the main barrier of zk-Creds multiple-credential linkage is the trust assigned by authorities to credentials issued by other parties. This is a largely political and social issue that lies outside the scope of this paper, but is worth mentioning in the context of inter-systems interactions.

The use of Groth16, in addition to the linkage proofs that enables interoperability, also provides other desirable properties. Like the RSS construction considered, Groth16 allows re-randomization with a weak form of simulation extractability. This guarantees that even if a malicious party has seen some proofs before, it cannot prove a new statement without knowing the witness. Given how frequent proofs may be required for decentralized identities, some form of information leak must be anticipated. With Groth16, should an adversary see old proofs they would still be unable to create forged proofs without the witness. This is a desirable feature to prevent the 'capture' of information which is possible in RSS as their verification still involves information extraction from a given VC. 

Given how crucial digital identities are, it is not surprising that the biggest drawback of zk-Creds is the trusted set-up procedure, followed by the practical integration of different credentials to take advantage of the linkage proofs enabled by Groth16. As mentioned, every application using Groth16 must have an initial trusted set-up phase which would need to be conducted after issuing a VC for every user. In addition to keeping the toxic waste of a CRS, any small errors in the whole protocol as demonstrated by an incorrect input aliasing can lead to falsifiable proofs which would be fatal to the implementation of a trust-worthy digital identity system. For instance, security audit company Beosin discovered that the zk-SNARK project Semaphore where the uin256 type represented a larger range of values than what is accepted in the arithmetic circuit, leading to proofs that can be used more than once and therefore enabled double spending \cite{Beosin}. There have been some notable improvements to the toxic waste to address these concerns. \cite{Bowe:2017} developed a distributed trusted set-up that is secure as long as at least one of the parties have honestly deleted their toxic waste. Though \cite{Bowe:2017} notes that there is an issue of robustness as all parties must be fixed before starting the CRS and must be active throughout the whole trusted set-up phase, this may not be a significant barrier for a VC depending on the implementation and connectivity. Further improvements such as \cite{Kohlweiss:2021} was able to avoid using the random beacon model to further simplify the process. Note that not all of the trusted set-up improvements are ready for deployment with some being in the theory phase so a more nuanced discussion will be needed. The issue of fatal errors in the set-up phase is a more practical one, and the best mitigation would be an open-source approach with consistent audits. However, it is possible to remove the trusted set-up entirely by replacing the ZK-SNARK component of zk-Creds with another protocol such as ZK-STARK which does away with this phase. The trade-off would be a significantly slower performance which may inhibit the system. 

The last drawback is that of witness generation. As mentioned in previous sections, zk-Creds uses witnesses to generate proofs by the prover. Witnesses in zk-Creds are specifically based on the credential issued list. However, the witness must always be generated from the most recent list of issued credential to ensure that a credential is still valid at the time of generating a proof. Though this does not seem to be problematic in theory, in practice it may make operations such as pre-processing difficult if there is a high volume of credentials being issued at a time from multiple issuers which requires the issuer list to be frequently updated. Users with multiple credentials would have to keep a witness per issuer on top of this. Though \cite{Rosenberg:2023} proposes some form of batching addition to potentially mitigate this issue, it still makes the capability to pre-compute proofs cumbersome. In addition, there are also privacy considerations for low-update and low-use systems. Generating a new witness may be matched to a credential as generating a witness requires showing a credential. This may be matched to a given proof if there is only a low number of witness generation and participants. Again, batching approach may mitigate this, but would be a context-specific deployment. Realistically, given the context of decentralized ID systems, this issue may be less applicable due to the target volume of users and frequency of anonymous disclosure. That being said, without the ability to pre-compute witnesses, it may not be the worst aspect as performance indicates zk-Cred proof generation will run similarly to RSS.

\subsection{Verdict}
Having considered the merits and drawbacks of each approach, we can conclude its suitability for decentralized identities. Clearly, there are aspects where zk-Creds provide advantage that RSS cannot replicate, and vice versa. In the case of zk-Creds, it is able to provide a much higher level of privacy as verifiers will never have to see the credential itself, allow more complex access criteria, and a superior performance compared to RSS. However, the issues of a trusted-setup and witness generation are factors that deter adoption. On the other hand, RSS has runs on a competitive speed and size when compared holistically against similar signatures with a customizable level of user privacy without the aforementioned issues facing zk-Creds. Yet, it cannot match zk-Creds in terms of the level privacy available and leaves the burden of information extraction for access criteria checks to the verifier. 

Having gathered concrete performance metrics in addition to considering the cryptographic features, we conclude that RSS is a much more convenient implementation for decentralized ID systems that should be adopted first. This does not mean that zk-Creds should be ruled out entirely. There are good use cases for them, but the current drawbacks indicate that this protocol is best adopted at a later stage. RSS is a more attractive scheme to adopt at an earlier stage of a decentralized ID system in part due to its ease to implement but also being able to offer an acceptable level of privacy. Allowing users the ability to selectively redact credential attributes aligns with the self-sovereign concept of decentralized identity by empowering users to select the level of privacy that they are comfortable with. This is still a partial solution for privacy, which is why in the following section we discuss the benefits of zk-Creds. But as a starting point, RSS performance metrics already compares well against signature schemes used in the decentralized identity space so we would not be looking at a radically different performance metrics. The RSS drawback of attribute verification needing to be computed by a verifier outside of the signature is also not necessarily an issue. At an earlier stage of development, it is likely that simpler access criteria will be more applicable such as proving one's age, one's country, etc. The computation cost imposed on such checks (range checks, string matching) is not costly and would be expected to add only a negligible overhead on the whole procedure. The RSS protocol can also be used to construct a mildly complex limited subscription credential as seen in \cite{Sanders:2021}, so more advanced uses of digital credentials is entirely possible with only RSS. 

However, the benefits of RSS can only extend so far. At a later or more mature stage of development, zk-Creds would be the more appropriate system. zk-Creds offers much better privacy by using proofs to avoid verifiers from seeing the actual credential itself. There is no need for credential holders to select which part of a VC they would like to redact, as the verifiers will never see any part of it in the first place but be provided with a verifiable proof that their access criteria is met. This is a much more complicated protocols (especially through the use of ZK-SNARKS which heavily involves pairing-based cryptography, arithmetic circuits, and homomorphic mappings) that will require much more awareness between stakeholders instead of the simpler concept RSS provides before adoption. Avoiding the implementation issues that trusted set-ups are liable to stringent checks will also take time, and a feedback process must be more intensive due to the possible cost of falsifiable proofs whereas RSS security rests on simpler digital signatures. 

As discussed in the previous section, the risk-prone trusted set-up mechanism can be made more secure through a distributed variant \cite{Bowe:2017}. Though some balance needs to be struck with regards to the number of participants in the interest of performance, a system that starts out with zk-Creds as the main verification protocol but few initial participants is at a higher risk of toxic waste not being deleted by any parties than one with much more participants involved at the trusted set-up phase happening at a later stage. A later stage adoption of zk-Creds would also be friendlier for witness generation. Given the expected increase in the number of participants, witness-batching is more feasible to preserve user anonymity, which is a possible privacy leak as mentioned in the previous section when there is not enough participants to obscure the witness request and proof verification processes. One may also be less liable to constant updates on the issued list of credentials depending on usage. For instance, a digital voter ID may see a smaller stream of users registering at a later stage instead of a persistent addition at the first stages of using this system. Some form of registration batching would also be possible depending on the exact numbers, which would bring down the number of updates to a credential list making witness updates less frequent. 

Adopting zk-Creds at a later stage is possible due to the Groth16 linkage proofs. Without this feature, one would have to start from scratch again when adopting a new protocol for verification which is a notable push-factor for improving identity systems. Linkage proofs, as demonstrated in \cite{Rosenberg:2023} using US NFC-enabled passports, allows parties to build on top of existing identity infrastructures without changing how they operate. It is therefore possible to initially adopt RSS as the basis for a decentralized ID verification but later transition to zk-Creds as parties look into working with different credentials and attributes that were not initially a part of the system with no material cost to modifying the system. This is not a feature that is possible with RSS, where verifiers would have to change their verification protocol to accept other verification methods used by different credentials. Crucially, linkage proofs also preserve privacy of other credentials and attributes, and would go a long way in enabling interoperability as other issuers would mitigate privacy concerns as public inputs can be hidden when forming proofs.

In short, given the performance metrics and cryptographic features of both protocols, we conclude that RSS is more appropriate to utilize during the earlier stages of an identity system whilst zk-Creds is better suited for a later-stage deployment which can be implemented without any changes to the initial system but rather to build on top of it. Additionally, a later-stage deployment of zk-Creds mitigates some issues regarding trusted set-ups and witness generation due to the number of participants and batching techniques available respectively whilst RSS provides user-enabled privacy at a competitive performance metric that can be implemented much earlier in decentralized identity systems without exposure to risks that zk-Creds are vulnerable to at an earlier stage.

\section{Conclusion}
This work has examined on two privacy-preserving protocols for verification in decentralized identity systems. RSS allows users to redact attributes of a credential, which enables users to select which attributes they would like to present to a verifier. zk-Creds is a protocol that uses ZK-SNARKS to present verifiers with proofs that shows an access criteria is met but without having to disclose their credential. Though both utilize pairing-based cryptography, they offer different benefits and drawbacks for decentralized identity systems.

Having implemented these two protocols on a sample VC, we find that RSS offers competitive performance in terms of speed and size when compared against other signature schemes currently being standardized for VCs. In addition, RSS empowers users by allowing them the flexibility to choose what level of privacy they would be comfortable with and can be used to build anonymous subscription-based access with applications in public transport. However, RSS still moves the burden of verifying access criteria to the verifier and would have difficulties when considering interoperability. We argue that these considerations makes RSS a good candidate for an early-stage ID system. It is able to offer user privacy, but the expected simpler access criteria would not impinge performance and interoperability considerations may done at a later stage. 

Here is where zk-Creds come into play. Through ZK-SNARKs and linkage proofs, it is able to operate with different credentials and does not compromise on privacy. Verifiers will never see the actual credential of a user but can set complex access criteria. In addition, zk-Creds does not need any signing mechanism as the verification relies on a witness based on a path on a Merkle tree. There are concerns about the trusted set-up requirement of zk-Creds and witness generation which may accidentally leak a user's identity. These factors make zk-Creds better suited for a later-stage adoption where interoperability, maximum privacy, and complex access-criteria can be handled by the protocol. The issues of trusted set-ups and witness generation are also partially mitigated with a later-stage implementation as a distributed variant with more participants lessens the risk of toxic waste being kept and less frequent sign-ups can lead to batching to remove the need to constantly update the witness. Thus, both techniques need not exist in opposition, but would be synergistic when implemented at different stages of a decentralized identity system.

Though our conclusion presents a cohesive application for these two techniques, there are several areas of RSS and ZKPs that merit future investigation. One direction of exploration is to examine any post-quantum variants for the two techniques. The US National Institute of Standards and Technology (NIST) has recently wrapped up its first competition of quantum-resistant cryptography. In this competition, lattice-based and hash-based protocols have been selected but future rounds are being conducted to expand the range of mathematical structures utilized by considering code-based cryptography.

Currently, we know of a lattice-based and a hash-based variant of the RSS \cite{Zhao:2022} \cite{Zhu:2022}. Even within these standards, there are variants of the structure that can provide differences. For instance, Learning with Errors is a popular hardness problem used to create lattice-based protocols. But, there are also variants that constructs Learning with Error over a Ring and a Module \cite{Amit:2019}. There are also other structures that may be quantum-secure such as isogenies. Within ZK-SNARKS, there is a lattice-based variant \cite{Albrecht:2022}. Again, it is worth investigating alternative constructions to ensure preparedness against potential quantum-adversaries that may break the current security assumptions of RSS and ZK-SNARKs. 

\section{Acknowledgments}
This work was supported, in whole or in part, by the Bill and Melinda Gates Foundation [INV-001309]. Under the grant conditions of the Foundation, a Creative Commons Attribution 4.0 Generic License has already been assigned to the Author Accepted Manuscript version that might arise from this submission.

\printbibliography
\end{multicols}
\end{document}